# Suffix Stripping Problem as an Optimization Problem

B. P. Pande, Pawan Tamta, H.S. Dhami

**Abstract**

Stemming or suffix stripping, an important part of the modern Information Retrieval systems, is to find the root word (stem) out of a given cluster of words. Existing algorithms targeting this problem have been developed in a haphazard manner. In this work, we model this problem as an optimization problem. An Integer Program is being developed to overcome the shortcomings of the existing approaches. The sample results of the proposed method are also being compared with an established technique in the field for English language. An AMPL code for the same IP has also been given.

## 1. Introduction

Suffix stripping is an important tool in the toolbox of Information Retrieval (IR) systems. In simple words, the problem can be expressed as: Given a word say TALKLESS, we have to remove word endings to get the stem word, TALK. This stem must be the root of all the inflected or variant forms of TALK such as TALKS, TALKING and TALKED. In this example TALK is the stem as it is the root of above inflected forms. Stemming or conflation is the process of finding the invariant string of a word that represents the root of the word. This is an active tool in IR environment to conflate inflected words that may be related to same meaning. Stemming can enhance the retrieval effectiveness, research has shown that it enhances the recall [11]. Stemming algorithms have been studied in computer science since the 1960s. The first published stemmer was coined by Julie Beth Lovins in 1968 [13]. Stemmers are very common elements in query systems, such as Information Retrieval systems, Web search engines and so on. Google search adopted word stemming in 2003. Examples of products using stemming algorithms are search engines such as Lycos and Google, and also thesauruses and other products using Natural Language

Processing for the purpose of IR. Stemming is used to determine domain vocabularies in domain analysis [7]. Many commercial companies have been using stemming since the 1980s and have produced lexical and algorithmic stemmers in many languages. The Snowball stemmers [20] have also been compared with such commercial lexical stemmers with varying results [22].

There is a rich literature of conflation algorithms. Automatic implementation of stemming algorithm may either depends on a pure linguistic approach, where prior morphological knowledge of a particular language is needed, or employs statistical techniques based on some statistical principles that may infer word formation rules for a targeted language. The linguistic techniques generally apply a set of transformation rules and endeavor to cut off known affixes. It is evident that these rules have to be carried out by the linguistic experts in a particular language which incurs, obviously, a manual labor. Some highlighted language dependent conflation algorithms are Lovin's algorithm [13], Dawson's algorithm [5], Porter algorithm [19], Paise/Husk algorithm [17], Krovetz algorithm [12], Harmans's 'S' stemmer [10] etc. Some recent trends can also be seen towards rule based stemming [1], [6], [18], [21].

Another limitation of linguistic techniques is their inability to work in a multilingual environment. Statistical techniques endeavor to cope up with this and allow dealing with new languages in the system even if little or no linguistic knowledge is available thereby ensuring minimum effort. This can be a crucial advantage especially for IR purpose where search is triggered for documents written in different languages. Successor variety word segmentation by Hafer and Weiss [9], Corpus based stemming by Jinxi Xu and W. Bruce [23], N-gram stemming by James Mayfield and Paul McNamee [15], hidden markov model (HMM) based automatic stemmer by Massimo Melucci and Nicola Orio [16], YASS stemmer by Prasenjit Majumder, et al [14], Context sensitive stemming by Peng Funchun et al. [8] are some of the important literature articles under statistical domain of stemming.

The principal urge is the development of the algorithms to tighten the gap between two categories. In this paper, we represent the suffix stripping problem as an optimization problem to fulfill the gap between two categories. We decided to target the multilingual environment. We address the problem as an optimization problem of maximization and frame an Integer Programming solution for it.

In section 2, basic definitions are being given. Section 3 covers the method with the development of the IP for suffix stripping. In Section 4, application of our technique is being shown over a set of word clusters. Section 5 highlights the conclusions and future work. In appendices, A.1 gives a comparison of outputs of our technique and Porter's Snowball stems [20] and in appendix A.2, an AMPL code for the proposed IP is given.

## 2. Preliminaries and Definitions

The process of conflation doesn't guarantee an output stem to be correct from linguistic point of view. Let us hypothesize a variable N-gram model where next character or symbol of a word is predicted using preceding symbols ($2 \leq N \leq$ word length). We are interested in forming bins of variable lengths where each bin contains symbol sequences scanned most recent. More particularly, we denote a given word $W$ of length $N$ as

$$W = w_1 w_2 w_3 \ldots w_i \ldots w_N$$

Where $w_i$ is any character, i.e. $w_i \in A$, set of alphabets. We define the frequency $f(w_1 w_2 \ldots w_i)$ as number of times a character sequence $w_1 w_2 \ldots w_i$ appears in a training corpus. We are interested in relative frequencies i.e. how many times a particular symbol $w_j$ follows the string $w_1 w_2 \ldots w_i$. These can be exploited as a probability estimate, maximum likelihood estimate (MLE). Mathematically, we define it as:

$$P_{MLE} = \frac{f(w_1 w_2 \ldots w_n)}{f(w_1 w_2 \ldots w_{n-1})} \qquad (2.1)$$

We've observed the following thing- a word in an inflectional language is composed of an essential stem and two arbitrary components: prefix and suffix. By arbitrary we mean that any random word may have both, at least one or no inflectional component. So, given a word, if we move ahead starting from the first character, we are supposed to cross an arbitrary sequence of prefix, followed by stem, followed by suffix. Thus, we can have three states, prefix, stem and suffix and there is/are transition(s) from one state to another in a word formation process. This assumption gives the idea of two arbitrary split points in a given word: one between prefix and stem, the other between stem and suffix. We calculate the MLE probability estimate for each next symbol in an input word, starting from symbol at $2^{nd}$ position. We infer if there is any relation between this estimate and partial word length (N-gram). From the test data, we observe that there is continuous increase in the value of probability estimate unless the split point occurs.

# 3. Formulation of the Algorithm

We emphasize to mention two objectives which an efficient algorithm must fulfill. First, the output obtained must be the root of all inflected words. Second, the algorithm must be language independent. A meaningful stem can be obtained by taking sequential probabilities into account. Before modeling the Integer Program for the problem, we discuss the following observations from the sample data:

1. In most of the cases, the highest probability is not unique across N-grams of a word.
2. Generally, the probabilities increase with the increase in the word length, a sudden decrease is subject to the transition from one state to another.
3. A continuous increase in the probabilities identifies the whole word as a stem.
4. Probabilities equal to 1 at last three or more positions may be associated with a common suffix, like *ing*.
5. Constant probabilities at the end for more than three positions may also indicate whole word as stem.
6. A continuous increasing sequence of probabilities at the end is associated with a consolidated ending, which can be removed to get a stem.

A general insight says to take an N-gram of a word as the stem which has the highest MLE score. But, as mentioned above, it is not unique. Moreover, sample results have shown that this method has yielded only 23 acceptable stems from a sample of 100 words. We therefore hypothesize how to incorporate all such observations in a single algorithm? While incorporating all these observations, we encounter randomness as the biggest problem. By considering all observations and objectives, we propose an Integer Program for suffix stripping problem. The strength of the algorithm lies in its capability to address all observations and to fulfill the objectives simultaneously. Moreover, since based on a mere probabilistic estimate, our Integer Programming technique is applicable to any language for which N-gram corpus frequencies are available.

## 3.1 Development of the Integer Program and Algorithm

For a given word *W* of length *n*, let $C_e$ (e=0, 1…n-1) be the MLE score of an N-gram, where $C_1$ corresponds to 2-gram, $C_2$ corresponds to 3-gram and so on. We assign a special MLE value of 0 to the first character of the input word, i.e. to the 1-gram, which means $C_0=0$. Let $\gamma_e$ be the binary variable associated with $C_e$, e=0, 1 … (n-1).

We define a variable $\gamma_N$ (where N=n-1) corresponding to the whole word as:

$$\gamma_N = \begin{cases} 1 & if\ C_N \geq C_{N-1} \\ 0 & otherwise \end{cases} \quad (3.1.1)$$

The integer program is given by

$$Maximize\ Z = \sum_{e=1}^{N-1} C_e \gamma_e + C_N \gamma_N \quad (3.1.2)$$

Subject to constraints

$$C_{e+1}\gamma_{e+1} - C_e\gamma_e \geq 0;\ e = 1 \ldots (N-2) \quad (3.1.3)$$

$$\gamma_e \in \{0,1\}, \forall e \quad (3.1.4)$$

Where $C_N\gamma_N$ in equation (3.1.2) is considered as a constant. Constraint (3.1.3) has been set to incorporate all six points observed above. It allows only those two consecutive characters as the part of a stem for which we have an increasing sequence of probabilities. The objective function considers the sequence of maximum length. By this idea, as a solution to the integer program, we get the sequence of $\gamma$ variables having value 1. The $\gamma$ variables which are not a part of the solution are set to value 0. We get the sequence of zero and non zero variables by this Integer Program. We pick that sequence which is larger (either of 0s or 1s) and the partial word string (N-gram) corresponding to the last entry of this sequence is taken as the stem. In case of ties, we prefer to take the latter sequence, i.e. longer word string or N-gram.

## 4. Experimental Results

To test our approach empirically, we first need sequential frequencies of N-grams from a rich corpus. For English language, we relied on COCA [4]. This corpus has two benefits; first it has a rich collection of English words and second, it's easy to calculate the N-gram frequencies using wild card [*]. Let's comprehend the working of our approach with the manifestation of the proposed IP. We consider the English word *Parsons* as a testing candidate. The sequential frequencies from COCA and probabilities are being tabled as under:

Table: 4.1: COCA frequencies for the word *Parsons*

| Word | N-Gram | Frequency (f) | $P_{MLE}$ ($C_e$) |
|---|---|---|---|
| Parsons | | | |
| | P | 1863235 | 0 |
| | Pa | 536621 | .288 |
| | Par | 250520 | .466 |
| | Pars | 2284 | .009 |
| | Parso | 606 | .265 |
| | Parson | 606 | 1 |
| | Parsons | 542 | .894 |

From equation 3.1.2, this can be formulated as:

$$\text{Max } Z = .288\gamma_1 + .466\gamma_2 + .009\gamma_3 + .265\gamma_4 + \gamma_5 + .894\times 0$$

Subject to the constraints

$$.466\gamma_2 - .288\gamma_1 \geq 0$$
$$.009\gamma_3 - .466\gamma_2 \geq 0$$
$$.265\gamma_4 - .009\gamma_3 \geq 0$$
$$\gamma_5 - .265\gamma_4 \geq 0$$

As we can see, $C_6 < C_5$ here, so by equation (3.1.1) we have taken $\gamma_6 = 0$. On solving this IP, we get the values of the variables as: $\gamma_1 = \gamma_2 = 0$ and $\gamma_3 = \gamma_4 = \gamma_5 = 1$. Sequence of non-zero variables i.e. 1s is longer than that of 0s, therefore the last variable is identified as $\gamma_5$. The stem is the word sequence (N-gram) associated with $\gamma_5$, which is *Parson*.

Let's apply the IP technique over some other language, consider the word *Dificilmente* (Portuguese). The sequential frequencies, taken from the corpus Corpus Do Português [3], are given as under:

Table: 4.2: Corpus Do Português frequencies for the word *Dificilmente*

| Word | N-Gram | Frequency (f) | $P_{MLE}$ ($C_e$) |
|---|---|---|---|
| Dificilmente | | | |
| | D | 737348 | 0 |
| | Di | 62719 | .085 |
| | Dif | 5714 | .091 |
| | Difi | 1639 | .286 |
| | Dific | 1639 | 1 |
| | Difici | 190 | .115 |
| | Dificil | 190 | 1 |

|   | Dificilm | 178 | .936 |
|---|---|---|---|
|   | Dificilme | 178 | 1 |
|   | Dificilmen | 178 | 1 |
|   | Dificilment | 178 | 1 |
|   | Dificilmente | 178 | 1 |

Here $\gamma_{11}=1$ and on solving the IP, we get $\gamma_1=\gamma_2=\gamma_3=\gamma_4=\gamma_5=\gamma_6=0$ and $\gamma_7=\gamma_8=\gamma_9=\gamma_{10}=1$. Again, sequence of 0s is longer than that of 1s, the last variable is identifies as $\gamma_6$ and the N-gram corresponding to $\gamma_6$ is taken as stem viz. *Dificil*.

Underneath we are giving a table exhibiting the stems resulted with our IP based conflation technique. Five different clusters are being taken into consideration. First five clusters are of English words, and the sixth is of Spanish words. For latter set of words, the sequential frequencies are taken from the corpus Corpus Del Español [2].

Table: 4.3: Clusters and their IP stems

| Cluster | Word | Stem | Cluster | Word | Stem |
|---|---|---|---|---|---|
| 1 | Create | Creat | 4 | Change | Change |
|   | Creates | Creat |   | Changes | Change |
|   | Created | Creat |   | Changed | Change |
|   | Creating | Creat |   | Changing | Chang |
|   | Creative | Creat |   | Changeable | Change |
| 2 | Include | Includ | 5 | Complete | Complete |
|   | Includes | Includ |   | Completes | Complete |
|   | Including | Includ |   | Completed | Complete |
|   | Included | Includ |   | Completing | Complet |
| 3 | Announce | Announc | 6 | Trabajan | Trabaj |
|   | Announces | Announc |   | Trabajar | Trabaj |
|   | Announced | Announc |   | Trabajado | Trabaj |
|   | Announcing | Announc |   | Trabajador | Trabaj |

## 5. Conclusions and Future work

The efficiency of current work is being concluded on the basis of the output obtained for 100 randomly chosen words, and comparing them with outputs of Porter's Snowball stemmer [20] (Appendix A.1). For English language, Porter's stemmer is being used very widely and has become the de facto standard algorithm. The survey has been categorized in three sections, first section contains the results identical to that of Porter (58), second section illustrates the morphologically better results (18) and in third section, we put the results close/inferior to that of Porter (24).

We endeavored to develop a probabilistic stemmer which should be language independent and came out with one which is dependent on mere character frequencies rather than morphological knowledge. This equips our technique to survive in a multilingual environment where need for specific morphological knowledge has been lessened. The notion of sequential probability estimate has been coined and for the first time in literature, the conflation problem is redefined as an optimization problem for which an Integer Program based solution is devised by us.

This work can have a two folded future prospects, one is the measurement of the hardness of the problem. A linear programming relaxation and development of some inequalities may result in some polynomial bounds. Second the development of new facet thereby defining inequalities that may enhance its efficiency and may put it more close to the morphological stem.

## 6. References

1. Araujo Lourdes, Zaragoza Hugo, Pérez-Agüera Jose R., Pérez-Iglesias Joaquín: Structure of morphologically expanded queries: A genetic algorithm approach, Data & Knowledge Engineering, Volume 69, Issue 3, 279-289 (2010).

2. Corpus Del Español. Available at < http://www.corpusdelespanol.org/>, visited 21 December 2013.

3. Corpus Do Português. Available at <http://www.corpusdoportugues.org/>, visited 21 December 2013.

4. Corpus of Contemporary American English (COCA). Available at <http://corpus.byu.edu/coca/>, visited 21 December 2013.

5. Dawson John: Suffix removal and word conflation. ALLC Bulletin, Volume 2, No. 3, 33-46 (1974).

6. English Joshua S.: English Stemming Algorithm, Pragmatic Solutions, Inc., 1-3 (2005).

7. Frakes, W. et al.: DARE: Domain Analysis and Reuse Environment, Annals of Software Engineering (5). 125-141 (1998).

8. Funchun Peng, Nawaaz Ahmed, Xin Li and Yumao Lu.: Context sensitive stemming for web search. Proceedings of the 30th annual international ACM SIGIR conference on Research and development in information retrieval, 639-646 (2007).

9. Hafer M. and S. Weiss: Word Segmentation by Letter Successor Varieties, Information Storage and Retrieval, 10, 371-85 (1974).

10. Harman Donna: How effective is suffixing? Journal of the American Society for Information Science, 42, 7-15 (1991).

11. Kraaij Wessel and Pohlmann Renee: Viewing stemming as recall enhancement. Proceedings of the 19th annual international ACM SIGIR conference on Research and development in information retrieval, 40-48 (1996).

12. Krovetz Robert: Viewing morphology as an inference process. Proceedings of the 16th annual international ACM SIGIR conference on Research and development in information retrieval, 191-202 (1993).

13. Lovins, J. B.: Development of a stemming algorithm. Mechanical Translation and Computational Linguistics, 11, 22-31 (1968).

14. Majumder Prasenjit, et al.: YASS: Yet another suffix stripper. ACM Transactions on Information Systems. 25(4), Article No. 18 (2007).

15. Mayfield James and McNamee Paul: Single N-gram stemming. Proceedings of the 26th annual international ACM SIGIR conference on Research and development in information retrieval, 415-416 (2003).

16. Melucci Massimo and Orio Nicola: Design, Implementation, and Evaluation of a Methodology for Automatic Stemmer Generation. Journal of the American Society for Information Science and Technology, 58(5), 673–686 (2007).

17. Paice Chris D.: Another stemmer. ACM SIGIR Forum, 24(3). 56-61 (1990).

18. Pande B. P. and Dhami H. S.: Application of Natural Language Processing Tools in Stemming. International Journal of Computer Applications, 27(6):14-19 (2011).

19. Porter M. F.: An algorithm for suffix stripping. Program 14, 130-137 (1980).

20. Porter, M. F.: Snowball: A language for stemming algorithms (2001). Available at <http://snowball.tartarus.org/>, visited 22 December 2013.

21. Tamah Eiman , Shammari-Al.: Towards an error free stemming, IADIS European Conference Data Mining, 160-163 (2008).

22. Tomlinson Stephen: Lexical and Algorithmic Stemming Compared for 9 European Languages with Hummingbird SearchServer$^{TM}$ at CLEF 2003. CLEF 2003: 286-300 (2003).

23. Xu Jinxi and Croft Bruce W.: Corpus-based stemming using co-occurrence of word variants, ACM Transactions on Information Systems. Volume 16 (1)1, 61-81 (1998).

# Appendices

**A.1:** Outputs of IP stemmer and Porter stemmer (Snowball) over 100 randomly chosen English words:

| 1: Outputs which are identical to Porter's | | | |
|---|---|---|---|
| *S. No.* | *Word* | *IP stem* | *Porter (Snowball) stem* |
| 1 | Parsons | Parson | Parson |
| 2 | Dilution | Dilut | Dilut |
| 3 | Agreement | Agreement | Agreement |
| 4 | Passion | Passion | Passion |
| 5 | Cutter | Cutter | Cutter |
| 6 | Museums | Museum | Museum |
| 7 | Fleet | Fleet | Fleet |
| 8 | Haze | Haze | Haze |
| 9 | Manace | Menac | Menac |
| 10 | Training | Train | Train |
| 11 | Ordaining | Ordain | Ordain |
| 12 | Abject | Abject | Abject |
| 13 | Overseas | Oversea | Oversea |
| 14 | Predicted | Predict | Predict |
| 15 | Lyric | Lyric | Lyric |
| 16 | Admonishing | Admonish | Admonish |
| 17 | Quiet | Quiet | Quiet |
| 18 | Unloaded | Unload | Unload |
| 19 | Alto | Alto | Alto |
| 20 | Casual | Casual | Casual |
| 21 | Admiring | Admir | Admir |
| 22 | Believing | Believ | Believ |
| 23 | Borrowed | Borrow | Borrow |
| 24 | Borrowing | Borrow | Borrow |
| 25 | Consulted | Consult | Consult |
| 26 | Consulting | Consult | Consult |
| 27 | Deceived | Deceiv | Deceiv |
| 28 | Deceiving | Deceiv | Deceiv |
| 29 | Employed | Employ | Employ |
| 30 | Employing | Employ | Employ |
| 31 | Explained | Explain | Explain |
| 32 | Explaining | Explain | Explain |
| 33 | Finished | Finish | Finish |
| 34 | Finishing | Finish | Finish |
| 35 | Gathered | Gather | Gather |
| 36 | Gathering | Gather | Gather |

| S. No. | Word | IP stem | Porter (Snowball) stem |
|---|---|---|---|
| 37 | Improved | Improv | Improv |
| 38 | Improving | Improv | Improv |
| 39 | Laughed | Laugh | Laugh |
| 40 | Laughing | Laugh | Laugh |
| 41 | Listened | Listen | Listen |
| 42 | Listening | Listen | Listen |
| 43 | Plucked | Pluck | Pluck |
| 44 | Plucking | Pluck | Pluck |
| 45 | Preached | Preach | Preach |
| 46 | Preaching | Preach | Preach |
| 47 | Enormously | Enorm | Enorm |
| 48 | Monthly | Month | Month |
| 49 | Solemnly | Solemn | Solemn |
| 50 | Frightfully | Fright | Fright |
| 51 | Swiftly | Swift | Swift |
| 52 | Blissfully | Bliss | Bliss |
| 53 | Viciously | Vicious | Vicious |
| 54 | Hopelessly | Hopeless | Hopeless |
| 55 | Briskly | Brisk | Brisk |
| 56 | Anxiously | Anxious | Anxious |
| 57 | Inwardly | Inward | Inward |
| 58 | Miserable | Miser | Miser |

2: Outputs which are morphologically better than Porter's

| S. No. | Word | IP stem | Porter (Snowball) stem |
|---|---|---|---|
| 1 | Vacancy | Vacancy | Vacanc |
| 2 | Ninety | Nine | Nineti |
| 3 | Bloodier | Blood | Bloodier |
| 4 | Despotic | Despotic | Despot |
| 5 | Eleventh | Eleven | Eleventh |
| 6 | Macabre | Macabre | Macabr |
| 7 | Quizzical | Quizzical | Quizzic |
| 8 | Undeveloped | Undeveloped | Undevelop |
| 9 | Carrying | Carry | Carri |
| 10 | Forbidden | Forbid | Forbidden |
| 11 | Abnormally | Abnormal | Abnorm |
| 12 | Diligently | Diligent | Dilig |
| 13 | Jubilantly | Jubilant | Jubil |
| 14 | Thankfully | Thankful | Thank |
| 15 | Reluctantly | Reluctant | Reluct |
| 16 | Wonderfully | Wonderful | Wonder |
| 17 | Delightfully | Delightful | Delight |
| 18 | Obnoxiously | Obnoxious | Obnoxi |

3: Outputs which are close/ inferior than Porter's

| S. No. | Word | IP stem | Porter (Snowball) stem |
|---|---|---|---|
| 1 | Refill | Refi | Refill |
| 2 | Stallion | Stall | Stallion |
| 3 | Braid | Bra | Braid |
| 4 | Midwife | Midwife | Midwif |
| 5 | Prophet | Prop | Prophet |
| 6 | Laceration | Lacerat | Lacer |
| 7 | Librarian | Librar | Librarian |
| 8 | Mainstream | Mainstre | Mainstream |
| 9 | Substitute | Substitute | Substitut |
| 10 | Freehand | Freeh | Freehand |
| 11 | Recognized | Recogni | Recogn |
| 12 | Devastating | Devastat | Devast |
| 13 | Abused | Abuse | Abus |
| 14 | Abusing | Abus/Abusing | Abus |
| 15 | Admired | Admi | Admir |
| 16 | Believed | Belie | Believ |
| 17 | Carried | Carrie | Carri |
| 18 | Decorated | Decorat | Decor |
| 19 | Decorating | Decorati | Decor |
| 20 | Forbidding | Forbidd | Forbid |
| 21 | Mended | Men/Mende | Mend |
| 22 | Mending | Mending | Mend |
| 23 | Nipped | Nipp | Nip |
| 24 | Nipping | Nipping | Nip |

**A.2**: AMPL program for the proposed IP

```
#----------Integer Program for Suffix stripping------
#----------------------Model------------------------------
param n;
param N=n-1;

set I={1..N};
var Gama{I} binary;
param C{I};
param const;

maximize z: sum {e in 1..N-1}
        C[e]*Gama[e]+const;

subject to

difference{x in 1..N-2}:
        C[x+1]*Gama[x+1]-C[x]*Gama[x]>=0;

#------------Data: For the word PARSONS---------------

data;
param n:=7;
param C:=
1 .288
2 .466
3 .009
4 .265
5 1
6 .894
;
var Gama[6] 0;
param const:=0;

option solver cplex;
solve ;

display z, Gama;
```